\documentclass{PoS}
\usepackage{tikz}
\usepackage{amsmath}
\usepackage[utf8]{inputenc} 
\usepackage{graphicx}
\usepackage{subcaption}
\title{Exotic and Conventional Quarkonium Physics Prospects at Belle II \\ {\large{Search for a Partner State to the $X(3872)$ at the $D^{*0}\bar{D}^{*0}$ threshold}}}
\ShortTitle{Search for the $X(4014)$}

\author{\speaker{Klemens Lautenbach}$^{a}$ and Jens Sören Lange$^{a}$ \thanks{on behalf of the Belle II collaboration}\\
        \llap{$^a$}Justus-Liebig-Universität Giessen\\
        E-mail: \email{klemens.lautenbach@physik.uni-giessen.de}}

\abstract{
The Belle II experiment at the SuperKEKB asymmetric $e^{+}e^{-}$ collider at KEK, Tsukuba, Japan, is recording data since 2018.
The expected high luminosity of $\mathcal{L}=8\cdot 10^{35}~$cm$^{-2}$s$^{-1}$ enables searches for yet unobserved exotic hadronic 
states in B-meson decays. We describe the search for a partner state of the $X(3872)$ at the $D^{*}\bar{D}^{*0}$ threshold.
The reconstruction of charmed mesons is tested with experimental data corresponding to an integrated luminosity of $504~$pb$^{-1}$ 
collected at the energy in the center of mass of $\Upsilon(4S)$ in 2018, and $2.62~$fb$^{-1}$ in 2019.

}

\FullConference{
European Physical Society Conference on High Energy Physics - EPS-HEP2019 -\\
			10-17 July, 2019\\
			Ghent, Belgium}

\begin{document}

\section{Motivation}
We are searching for a partner state of the $X(3872)$ at the $D^{*0}\bar{D}^{*0}$ threshold. 
In particular, models which describe the X(3872) as a $D^{*0}\bar{D}^{*0}$ molecule, favor the existence 
of the so-called $X(4014)$. Historically predicted already by Törnqvist \cite{Tornqvist}, 
recently more refined properties were predicted \cite{Cincioglu}. 
The existence of the $X(4014)$ is very sensitive to any admixture of conventional charmonium state in the wavefunction. 
For the $X(3872)$, the mass of the $\chi_{c1}'$ is higher, for the $X(4014)$, the mass of the $\chi_{c2}'$ is lower, leading to repulsion 
if there is a charmonium admixture in the wave function. If the admixture is around $10 - 30\%$, the $X(4014)$ destabilizes
and vanishes from the spectrum \cite{Nieves}.

\section{The Belle II Detector}
The Belle II experiment is located at the asymmetric $e^{+}e^{-}$ collider SuperKEKB in Tsukuba, Japan \cite{superkekb}. 
It is operating at $\sqrt{s}=10.58~$GeV, corresponding to the $\Upsilon(4S)$ resonance which decays dominantly to 
$B^{0}\bar{B}^{0}$ or $B^{+}B^{-}$. Similar to its predecessor Belle, it is a multi-layer detector consisting of various sub-detectors allowing for tracking, 
particle identification and energy measurement capabilities assorted in a barrel-shaped geometry. 
The acceptance in the laboratory frame is $2\pi$ in the azimuthal and $17^{\circ}<\Theta < 150^{\circ}$ in the 
polar plane. At the innermost layer, Belle II features a Vertex detector system (VXD) made up of a 
two-layer DEPFET Pixel Detector (PXD) and a four-layer silicon strip Vertex Detector (SVD). The inner tracking 
devices are surrounded by the Central Drift Chamber (CDC), a wire chamber tracking detector. 
Particle identification is done by a quartz-plate Time of Propagation detector (TOP) and an Aerogel 
ring-imagine-Cherenkov detector (ARICH), placed at the outer layer of the CDC. The electromagnetic calorimeter, 
consisting of Thallium-doped Cesium Iodide (CsI(TI)), is placed outside the particle identification devices. 
For $K_{L}^{0}$ and muon identification, the $K_{L}^{0}$ and muon detector (KLM) is placed outside of the ECL. 
It consists of alternating iron plates with scintillators and resistive plate chambers for detection and is also 
responsible for returning the magnetic flux. For details about the Belle II detector performance, we refer to \cite{Forti}, \cite{Oki}.
 
\section{Simulation of an $X(3872)$ partner state at the $D^{*0}\bar{D}^{*0}$ threshold}
\begin{figure}
\begin{center}
\begin{tikzpicture}[scale=1.0, every node/.style={scale=1.0}]
        \node at (0, 4) {$\Upsilon(4S)\rightarrow B^{+}B^{-}$};
        \draw (0.4, 3.8) -- (0.4, 3.5);
        \draw[->] (0.4, 3.5) -- (1.3, 3.5);
        \node at (2.5, 3.5) {$X(4014)K^{+}$};
        \draw (2, 3.3) -- (2, 3.0);
        \draw[->] (2, 3.0) -- (3.8, 3.0);
        \node at (4.8, 3.0) {$D^{0}\bar{D}^{0}\pi^{0}\pi^{0}$};
        \draw (4.1, 2.7) -- (4.1, 0.5);
        \draw[->] (4.1, 2.5) -- (5.4, 2.5);
        \draw[->] (4.1, 2.0) -- (5.4, 2.0);
		\draw[->] (4.1, 1.5) -- (5.4, 1.5);
		\draw[->] (4.1, 1.0) -- (5.4, 1.0);
		\draw[->] (4.1, 0.5) -- (5.4, 0.5);
		
		\node at (6.1, 2.5) {$K^{-}\pi^{+}$};
        \node at (6.3, 2.0) {$K^{-}\pi^{+}\pi^{0}$};
        \node at (6.55, 1.5) {$K^{-}\pi^{+}\pi^{+}\pi^{-}$};
        \node at (6.1, 1.0) {$K^{-}K^{+}$};
        \node at (6.3, 0.5) {$K_{S}^{0}\pi^{+}\pi^{-}$};
  \end{tikzpicture}
\end{center}
\caption{Decay chain used for simulation of the $X(4014)$.}
\label{fig:dec_chain}
\end{figure}
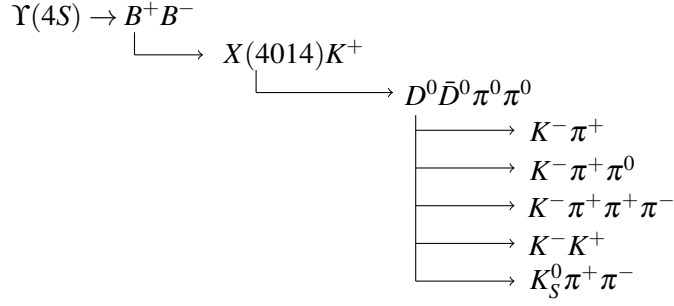

We simulate a potential, exotic charmonium resonance at the $D^{*0}\bar{D}^{*0}$ threshold assuming the same total 
 width as the $X(3872)$, $\Gamma < 1.2~$MeV \cite{Tanabashi}. To reconstruct the potential candidate, we use the 
$D^{*0}\rightarrow D^{0}\pi^{0}$ decays. Furthermore, we reconstruct the candidates directly in 
$D^{0}\bar{D^{0}}\pi^{0}\pi^{0}$, without requiring an intermediate resonance which results in wider 
phase space for the $D^{0}$ and $\pi^{0}$ momentum leading to a better separation of signal and background in the 
$X(4014)$ invariant mass spectrum. $D^{0}$ mesons are reconstructed in 
5 decay modes. The whole decay chain is summarized in Fig. \ref{fig:dec_chain}. 

\newpage
\subsection{Selection of Final State Particles}
We select only events with more than 5 tracks satisfying the criterion of having the probability of $\chi^{2}$ of the vertex fit 
when combining and fitting the reconstructed tracks $P(\chi^{2})>0.1\%$. 
$K^{\pm}$ and $\pi^{\mp}$ are reconstructed with a rather loose particle identification (PID) criterion of $>0.05$ and $>0.1$ respectively, 
the higher value in case of the $K^{\pm}$ is due to the higher purity of misidentified candidates up to $0.1$. 
The PID is defined as $\mathcal{L}_K/\Sigma \mathcal{L}_i$, where $\mathcal{L}_i$ is the likelihood for the particle being identified as electron, 
pion, deuteron, proton, muon or kaon. 
Using a higher PID decreases the overall reconstruction efficiency of the $B^{\pm}$ significantly while the increase 
in purity is negligible. For both particles, we select only those, for which the corresponding track originates within $|d0| < 0.5~$cm and 
$|dz|<3~$cm, where $d0$ defines the point of closest approach to the corresponding track and $dz$ being the distance in beam-direction with 
respect to the interaction point. Photons are selected by having at least 1.5 hits in the ECL crystals and  
energy greater than $25~$MeV. The fraction of deposited energy in $5\times 5$ over $3\times 3$ clusters should be greater than 0.9. \\

\begin{figure}
\begin{center}
	\includegraphics[width=.4\textwidth]{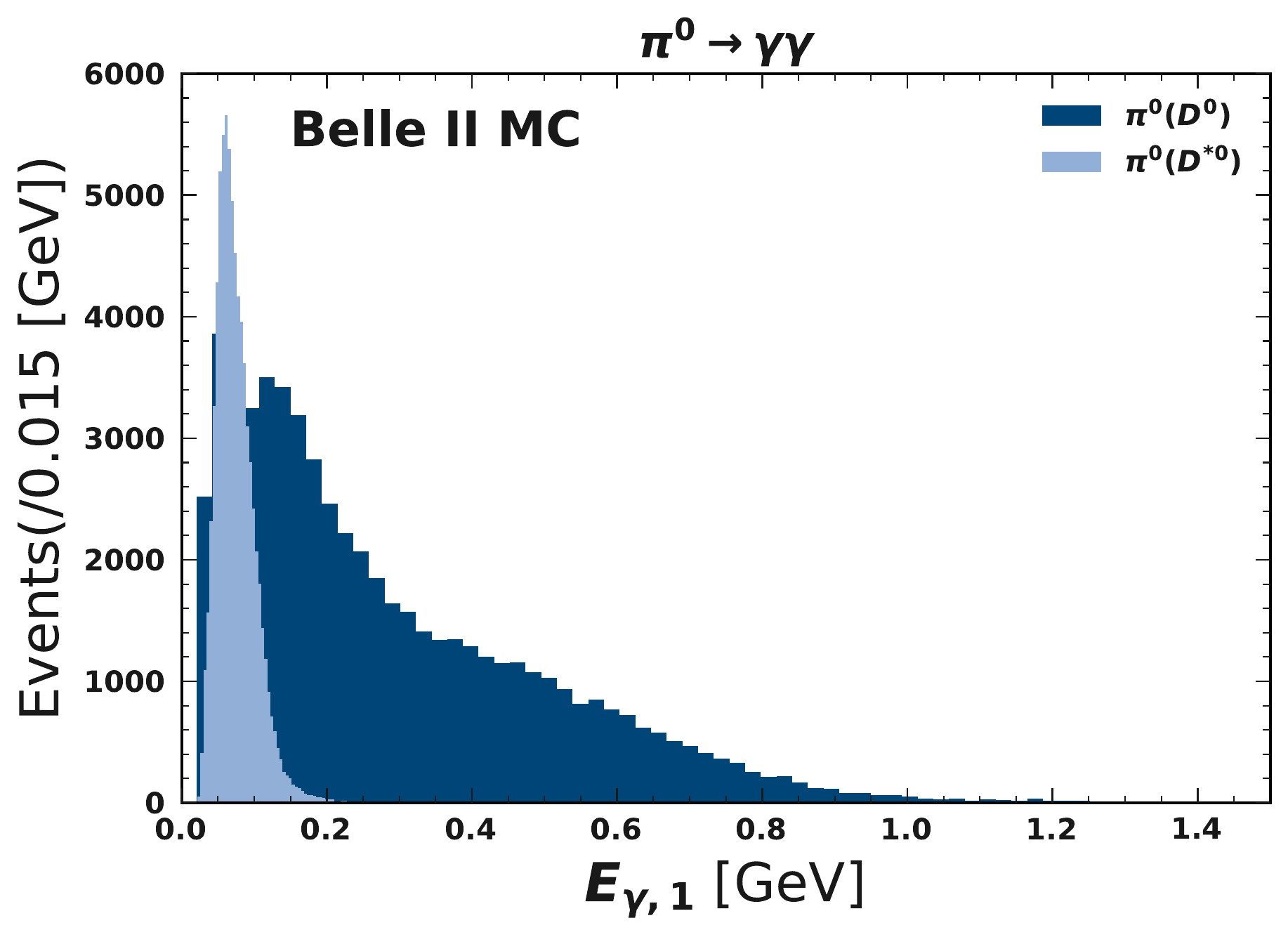}
	\includegraphics[width=.4\textwidth]{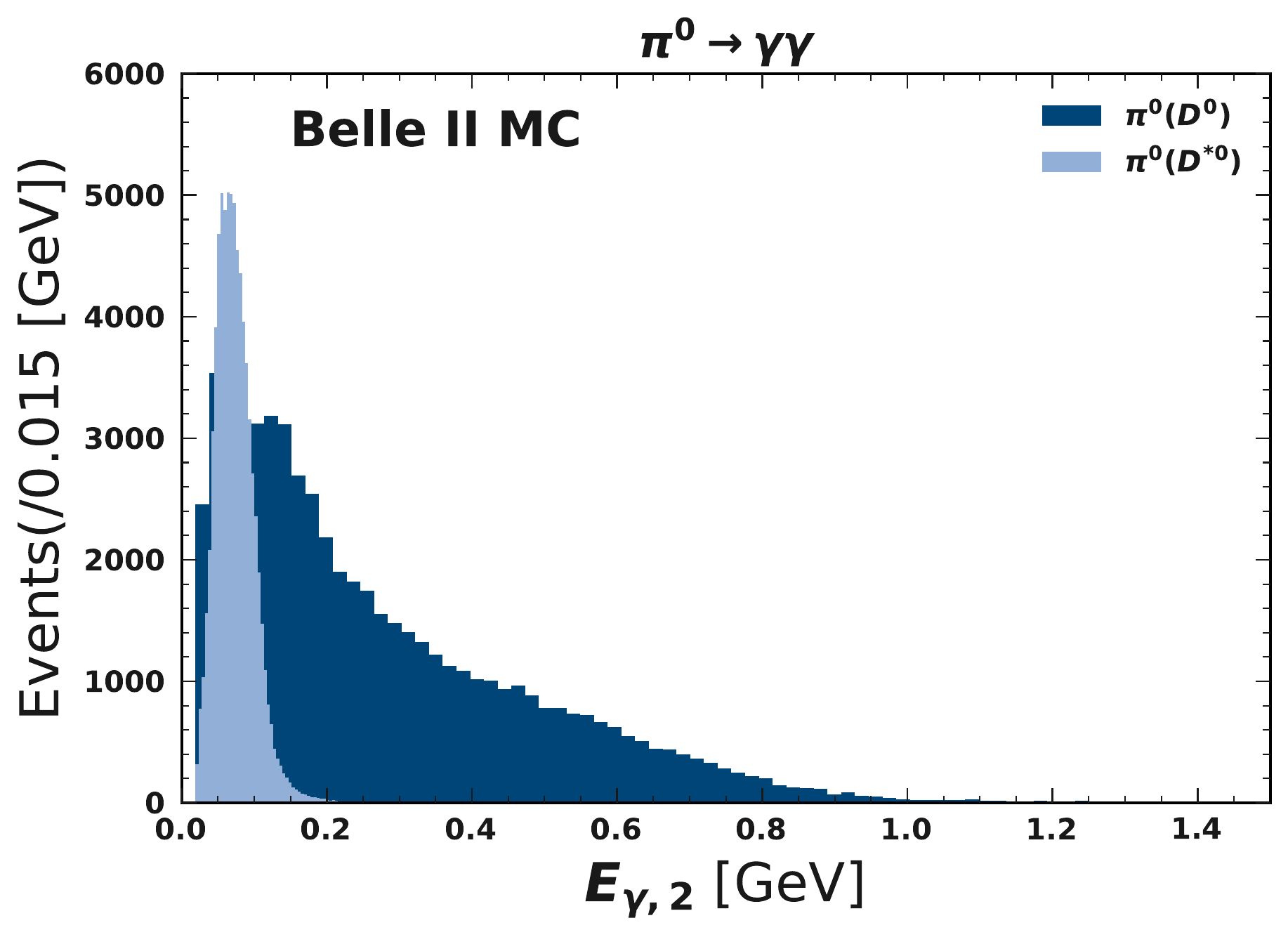}
\end{center}
\caption{Energy of the first (left) and second (right) photon of the $\pi^{0}$. The light blue color refers to $\pi^{0}$ from $D^{*0}$, in dark blue from $D^{0}$.}
\label{fig:pi0_sepa}
\end{figure}

$\pi^{0}$s originating from $D^{0}$ can populate a much broader momentum range as the ones from the $X(4014)$ decay. 
Since the $\pi^{0}$ is decaying into two photons, a higher momentum of the $\pi^{0}$ can be directly translated into higher photon energies 
as shown in Fig. \ref{fig:pi0_sepa}. Introducing a separation in the energy of $E_{\gamma}>0.12~$MeV for photons from $\pi^{0}$ stemming from $D^{0}$ and 
$E_{\gamma}<0.12~$MeV the ones originating directly from $X(4014)$, leads to a better reconstruction efficiency and, compared to a selection on the 
$\pi^{0}$ energy directly, to a significant gain in purity. Reconstructed $\pi^{0}$s are selected in a mass range of $0.11 < M < 0.16~$MeV/c$^{2}$ for 
candidates stemming from $D^{0}$ and $0.1<M<0.17~$MeV/c$^{2}$ for candidates from $X(4014)$. All candidates passing the above selection criteria are mass fit constrained.

\subsubsection{Reconstruction of Intermidiate $D^{0}$ States}
\begin{figure}
\begin{center}
	\includegraphics[width=.31\textwidth]{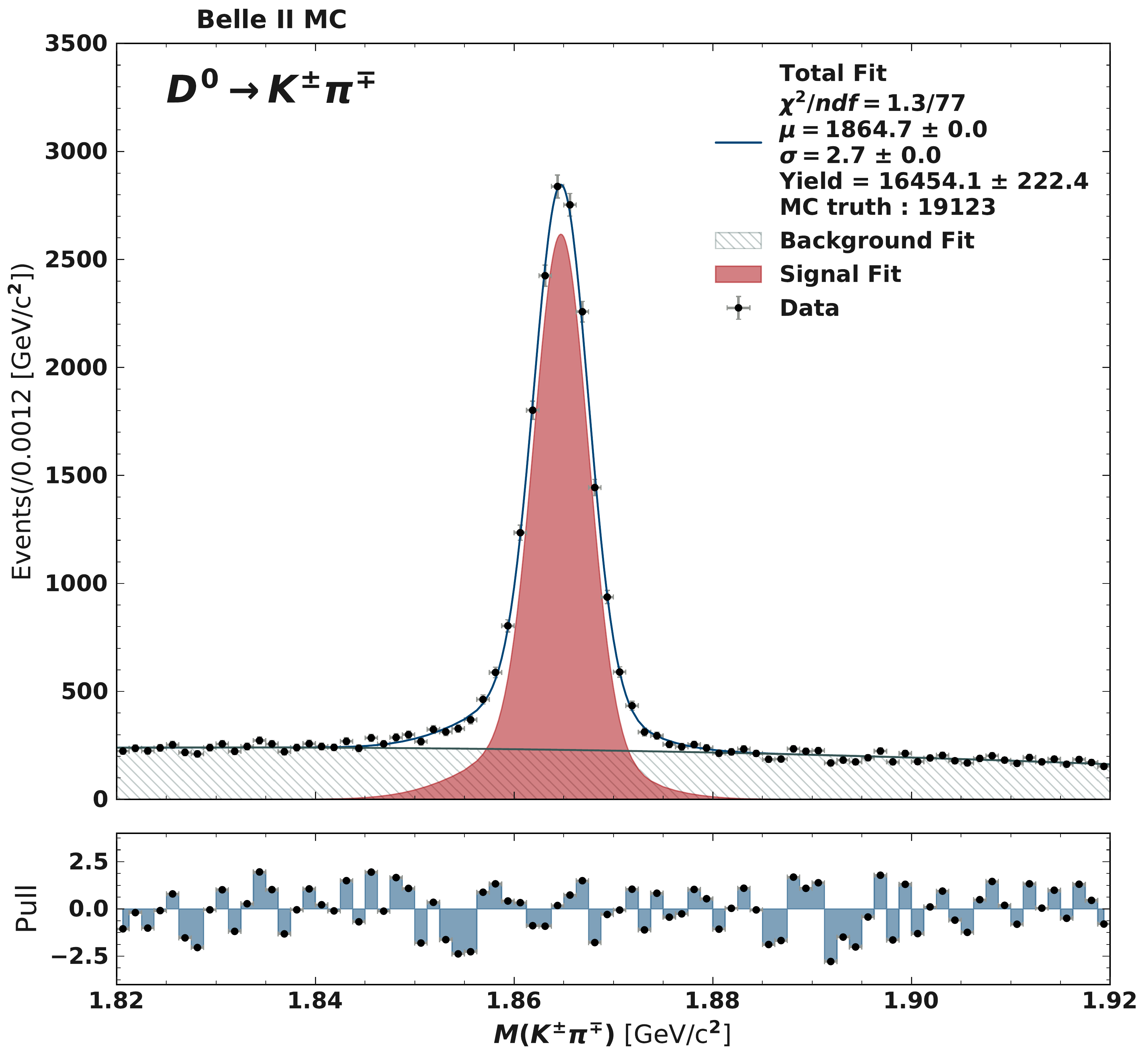}
	\includegraphics[width=.31\textwidth]{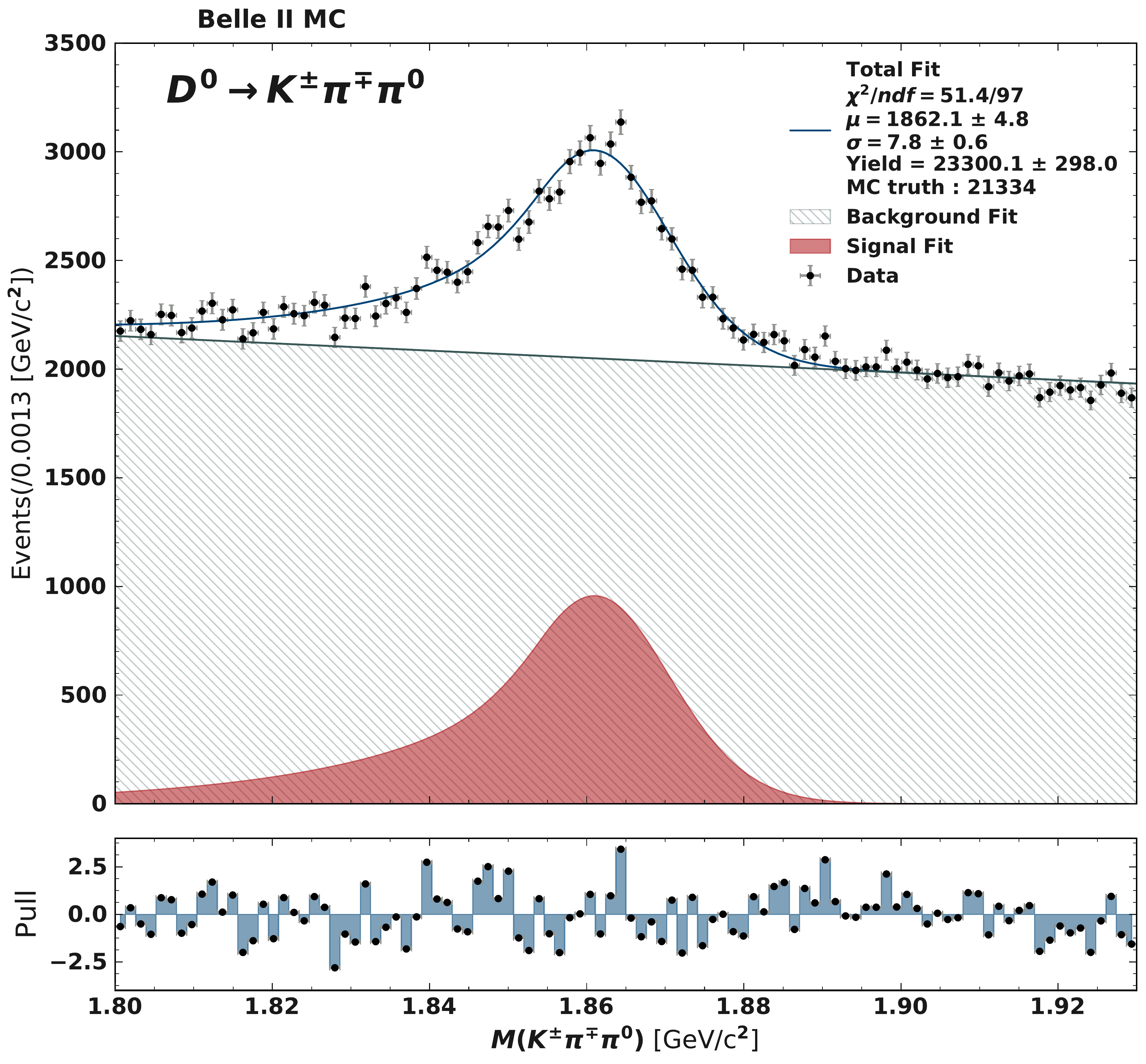}
	\includegraphics[width=.31\textwidth]{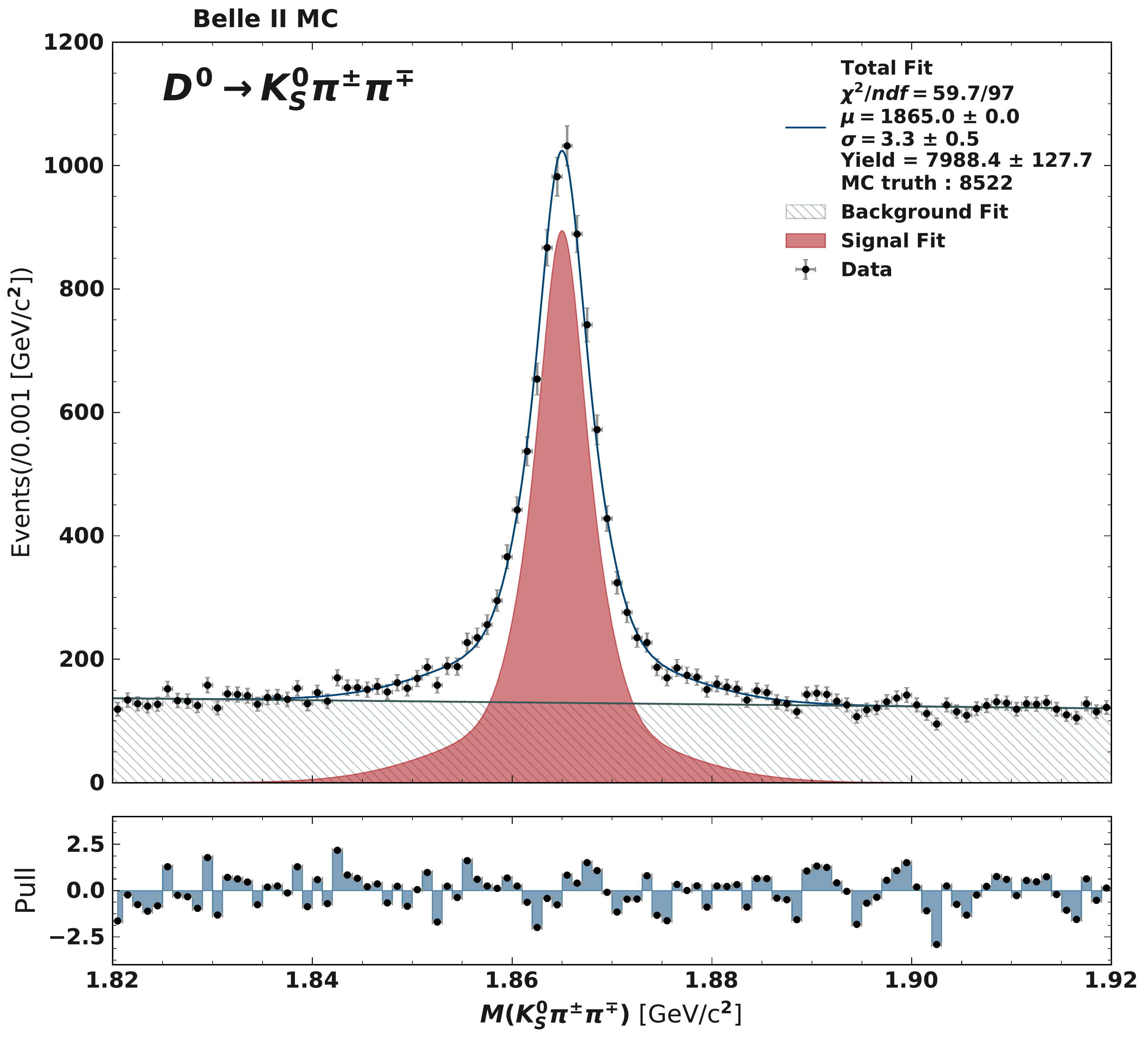}
\end{center}
\caption{Fitted $D^{0}$ mass peaks for a part of the decay channels shown in \ref{fig:dec_chain} in signal Monte-Carlo 
with 100000 events generated.}
\label{fig:D0_MC}
\end{figure}

As already shown before, we reconstruct the $D^{0}$ in 5 channels (Fig. \ref{fig:dec_chain}) which sum up to a total of 
$\approx 30\%$ of the $D^{0}$ branching fraction. In Fig. \ref{fig:D0_MC} we show the fitted invariant mass peaks for 
the different channels. 

We select all candidates within a $3\sigma$ region around the invariant mass mean and apply a vertex, then a mass fit constraint. 

\subsubsection{Reconstruction and Selection of $B^{\pm}$ Candidates}
We combine two $D^{0}$ and two $\pi^{0}$ to form a possible candidate for the $X(4014)$ resonance and select all candidates within a mass range of 
$3.9<M<4.06~$GeV/c$^{2}$. In order to reconstruct the $B^{\pm}$, a $K^{\pm}$ is added to the $X(4014)$ candidate and a preselection of $|\Delta E|<0.2~$GeV 
and $M_{BC}> 5.2~$GeV/c$^{2}$ is applied with  
$|\Delta E| = |E_{B}-E_{Beam}|$, the energy difference between the reconstructed $B^{\pm}$ and the beam and $M_{BC}=\sqrt{E_{Beam}^{2}-\Sigma p_{B}^{2}}$, 
the beam constraint mass.\\

After reconstruction, we apply a tighter selection window of $|\Delta E|<0.01~$GeV and $M_{BC}>5.27~$GeV/c$^{2}$ which removes a significant part of the combinatorial 
background due to wrong kinematics of the $B^{\pm}$ daughters. The purity of the sample increases  
from $0.6~\%$ to $3.9~\%$ while the effect on reconstruction efficiency is found to be negligible at the level of $0.2~\%$. 
Due to the high multiplicity of 9.4 $B^{\pm}$ candidates per event, we need to further improve our selection by applying a  
Best Candidate Selection (BCS) which is shown in Eq. \ref{eq:BCS}. 

\begin{equation}
\begin{split}
 \chi^{2}_{BCS} &= \Bigg(\frac{M_{D^{0}, 1}-M_{D^{0}}^{PDG}}{\sigma_{M_{D^{0}, 1}}}\Bigg)^{2} + \Bigg(\frac{M_{D^{0}, 2}-M_{D^{0}}^{PDG}}{\sigma_{M_{D^{0}, 2}}}\Bigg)^{2}                                          
 + \Bigg(\frac{M_{\pi^{0}, 1}-M_{\pi^{0}}^{PDG}}{\sigma_{\pi^{0}}}\Bigg)^{2} \\\nonumber
 &+ \Bigg(\frac{M_{\pi^{0}, 2}-M_{\pi^{0}}^{PDG}}{\sigma_{\pi^{0}}}\Bigg)^{2}
  + \Bigg(\frac{\Delta E}{\sigma_{\Delta E}}\Bigg)^{2} + \Bigg(\frac{M_{BC}-M_{B}^{PDG}}{\sigma_{M_{BC}-M_{B}^{PDG}}}\Bigg)^{2} + 
\Bigg[\frac{M_{\pi^{0}, D} - M_{\pi^{0}}^{PDG}}{\sigma_{\pi^{0}, D}}\Bigg]^{2} \\\nonumber 
 &+ (1 - \chi^{2}_{D^{0}_{i}}) + (1 - \chi^{2}_{K^{\pm}_{i}, \pi^{\pm}_{i}}) + (1 - kaonID_{i}) + (1 - pionID_{i})
\end{split}
\label{eq:BCS}
\end{equation}

\begin{itemize}
\item $M_{D^{0}, 1/2}$, $\sigma_{D^{0}, 1/2}$ : mass and fitted width (decay specific) of the first or second $D^{0}$
\item $M_{*}^{PDG}$ : nominal mass of the particle 
\item $M_{\pi^{0}, 1/2}$, $\sigma_{\pi^{0}}$ : mass and fitted width of the first or second $\pi^{0}$
\item $M_{\pi^{0}, D}$, $\sigma_{\pi^{0}, D}$ :  mass and fitted width of the $\pi^{0}$ in case of $D^{0}\rightarrow K^{\pm}\pi^{\mp}\pi^{0}$ decay
\item $|\Delta E|$, $\sigma_{\Delta E}$ : energy difference and fitted width of energy difference between $B^{\pm}$ and the beam  
\item $\sigma_{M_{BC}-M_{B}^{PDG}}$ : fitted width of the beam constrained mass minus the nominal B mass
\item $\chi^{2}_{D^{0}_{i}}, \chi^{2}_{K_{i}^{\pm}, \pi_{i}^{\pm}}$ : Goodness of fitted track
\item $kaonID_{i}, pionID_{i}$ : Kaon and Pion ID of the final state particles in the decay
\end{itemize}  

We calculate $\chi^{2}_{BCS}$ as shown in the equation above for each candidate and select the one with the smallest $\chi^{2}_{BCS}$ for each event. 
The purity of the signal sample increases from $3.9~\%$ to $21.4~\%$ while the reconstruction efficiency drops from $1.2~\%$ to $0.8~\%$, which is, 
due to the significant rise in purity, acceptable. 

\subsubsection{Estimation for discovery in an $2~$ab$^{-1}$ data set} 	
In order to understand whether an integrated luminosity equal to $\mathcal{L}=2~$ab$^{-1}$ is sufficient for the observation of the aforementioned $X(4014)$ at Belle II,
we estimate the number of 
signal events corresponding to a branching fraction of $\mathcal{B}=10^{-4}$ for the decay $B^{\pm}\rightarrow X(4014)K^{\pm}, X(4014)\rightarrow D^{0}\bar{D}^{0}\pi^{0}\pi^{0}$, 
similar to that of the $X(3872)$:\\

\begin{center}
$N_{evt}=\mathcal{L}_{int}\cdot \mathcal{B}\cdot 1.1~$nb$~\approx 200000$\\
\end{center}

Since we only take into account $D^{0}$ decays which sum up to $30\%$ of the total branching fraction, we do not consider the radiative $D^{*0}$ decay 
and we only reconstruct charged $B$-mesons, we expect an effective, detectable number of $15000$ signal events in $2~$ab$^{-1}$ of data. 
We use a convolution of a Breit-Wigner distribution with two Gaussians as the probability density function for the signal, a bifurcated Gaussian for the combinatorial 
background and $F_{genBKG}(x; p_{0}, p_{1}, p_{2}, x_{0}) = p_{0}(x-x_{0})^{p_{1}}\cdot e^{p_{2}(x-x_{0})}$ for the generic background.

\begin{figure}
\begin{center}
	\includegraphics[width=.5\textwidth]{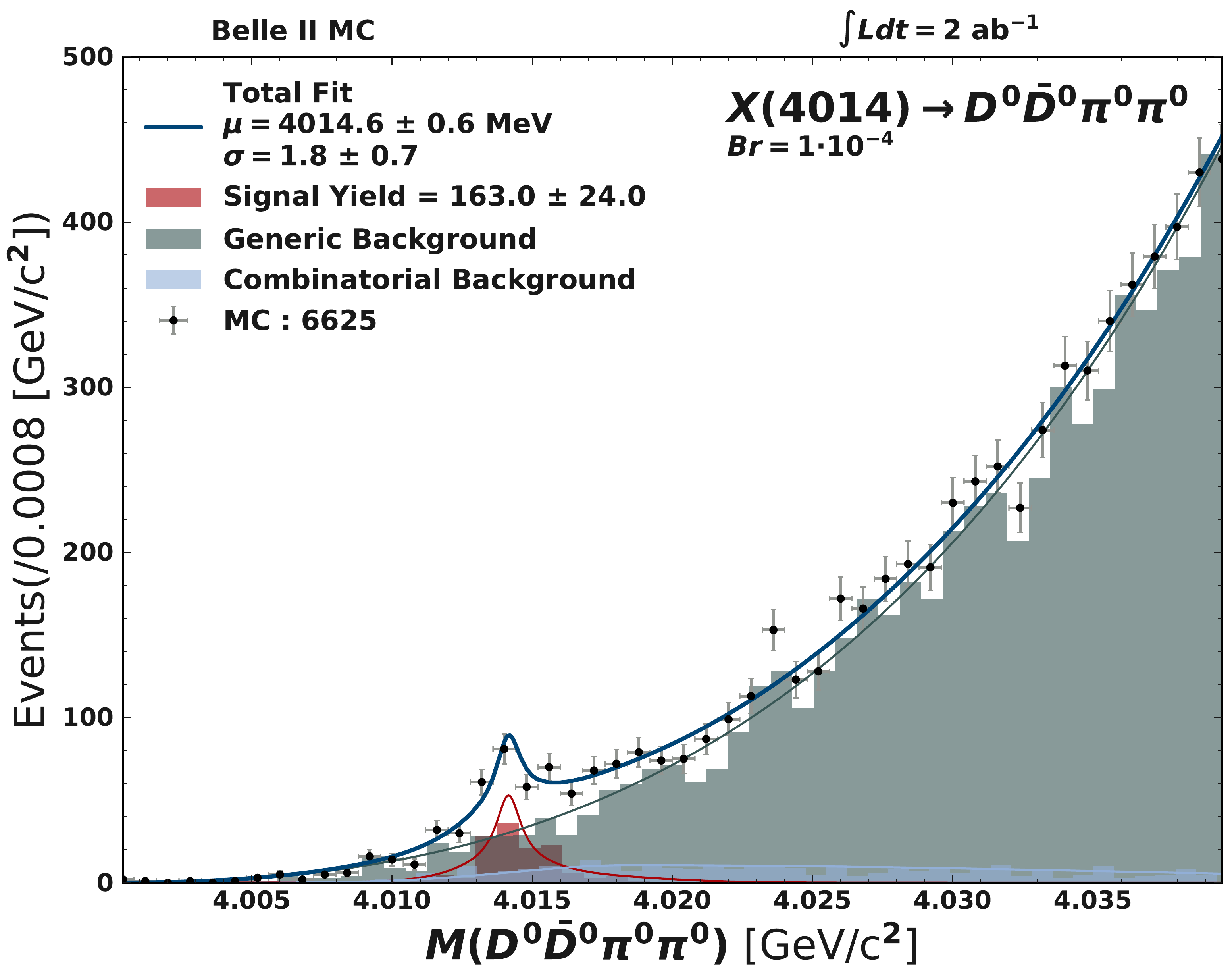}
\end{center}
\caption{Extended, unbinned maximum likelihood fit to the $X(4014)$ invariant simulated for a data set of $2~$ab$^{-1}$.}
\label{fig:Fit2ab}
\end{figure}

The total data set including all three components are then fitted with an unbinned maximum likelihood fit, using the sum of the functions above. The outcome can be seen in Fig. \ref{fig:Fit2ab}.
 
We perform the fit again with a model containing the two background component functions only, to get the likelihood for background only assumption. 
The significance is then calculated via $\sigma=\sqrt{2\cdot(ln(S+B)-ln(B))}$. If the $X(3872)$ partner exists and has a comparable branching fraction, 
we expect to see a peaking structure with $6.5~\sigma$ statistical significance.  

%
\section{$D$ Meson Reconstruction in Early Data}
Until Belle II collects more data, a full reconstruction of $B^{\pm}\rightarrow X(4014)K^{\pm}$ is not possible, but a first 
verification of the final state particle selections and fits on intermediate resonances might be feasible. In the following, we compare the two 
data sets collected so far during the commissioning run\footnote{Only 1/2 SVD and 2/32 PXD modules} 2018 ($0.504~$fb$^{-1}$) 
and the first full detector\footnote{Except PXD with 1/2 modules} run in 2019 ($2.62~$fb$^{-1}$).
\newpage
\begin{table}[h]
\begin{center}
\begin{tabular}{l|c|c}
2018 data, $0.504~$fb$^{-1}$ &  $\sigma$ [MeV] & Yield / fb$^{-1}$\\
\hline
$D^{0}\rightarrow K^{\pm}\pi^{\mp}$ 					& $4.1$	& $12748$\\
$D^{0}\rightarrow K^{\pm}\pi^{\mp}\pi^{0}$ 				& $9.9$	& $9484$\\
$D^{0}\rightarrow K^{\pm}\pi^{\mp}\pi^{\pm}\pi^{\mp}$ 	& $5.6$	& $1496$\\
$D^{0}\rightarrow K_{S}^{0}\pi^{\pm}\pi^{\mp}$ 			& $4.7$	& $556$\\
$D^{0}\rightarrow K^{\pm}K^{\mp}$ 						& $4.9$	& $1550$\\
\hline
2019 data, $2.62~$fb$^{-1}$ & & \\
\hline
$D^{0}\rightarrow K^{\pm}\pi^{\mp}$ 					& $3.2$	& $24771$\\
$D^{0}\rightarrow K^{\pm}\pi^{\mp}\pi^{0}$ 				& $9.7$	& $17111$\\
$D^{0}\rightarrow K^{\pm}\pi^{\mp}\pi^{\pm}\pi^{\mp}$ 	& $3.1$	& $5599$\\
$D^{0}\rightarrow K_{S}^{0}\pi^{\pm}\pi^{\mp}$ 			& $3.2$	& $3131$\\
$D^{0}\rightarrow K^{\pm}K^{\mp}$ 						& $2.9$	& $2794$ \\
\end{tabular}
\caption{Comparison of fitted width and yields for $0.504~$fb$^{-1}$ taken in 2018 and $2.62~$fb$^{-1}$ of 2019 data.}
\label{fig:signi}
\end{center}
\end{table}

\begin{figure}
    \centering
    \begin{subfigure}[b]{0.9\textwidth}
	\includegraphics[width=.31\textwidth]{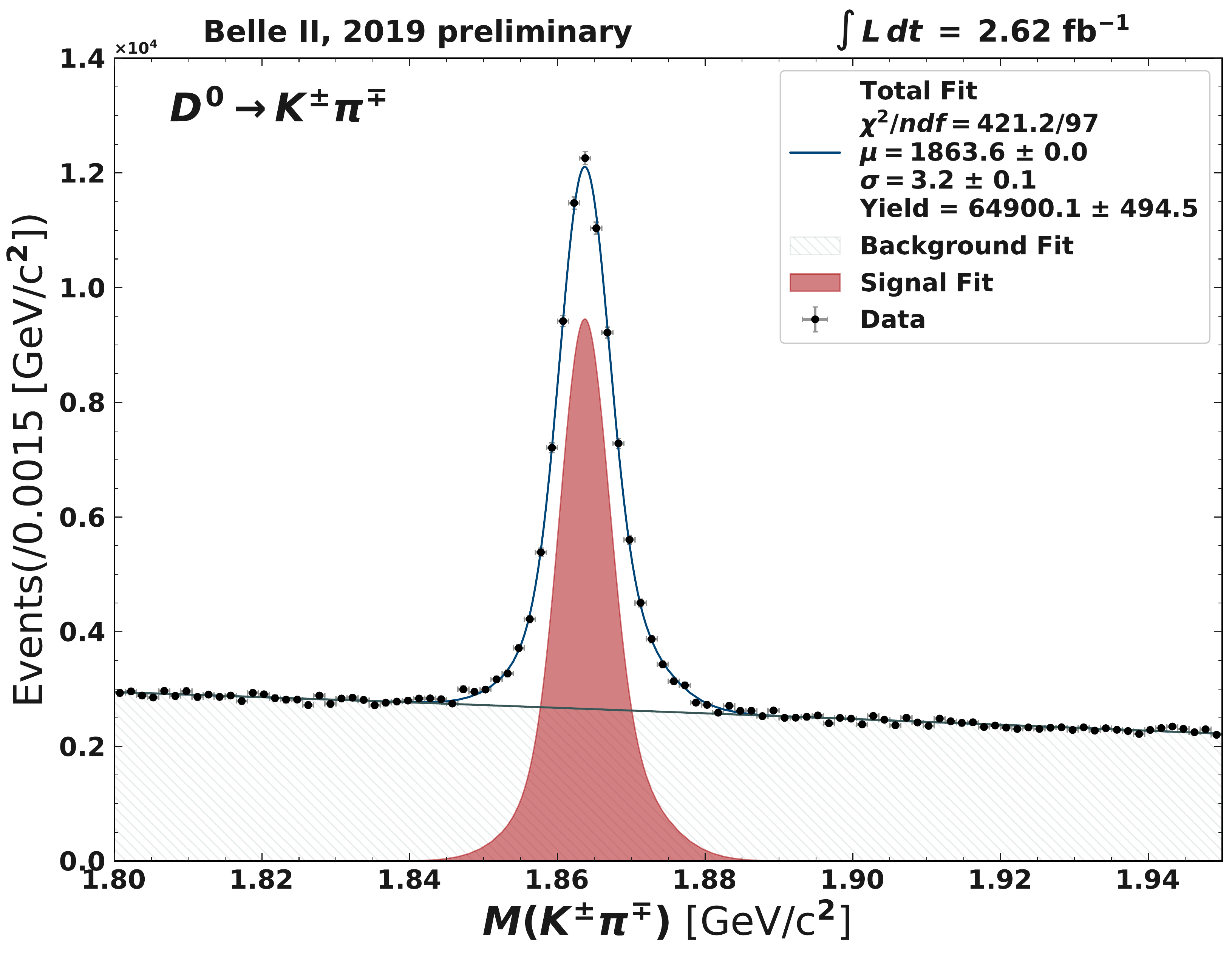}
	\includegraphics[width=.31\textwidth]{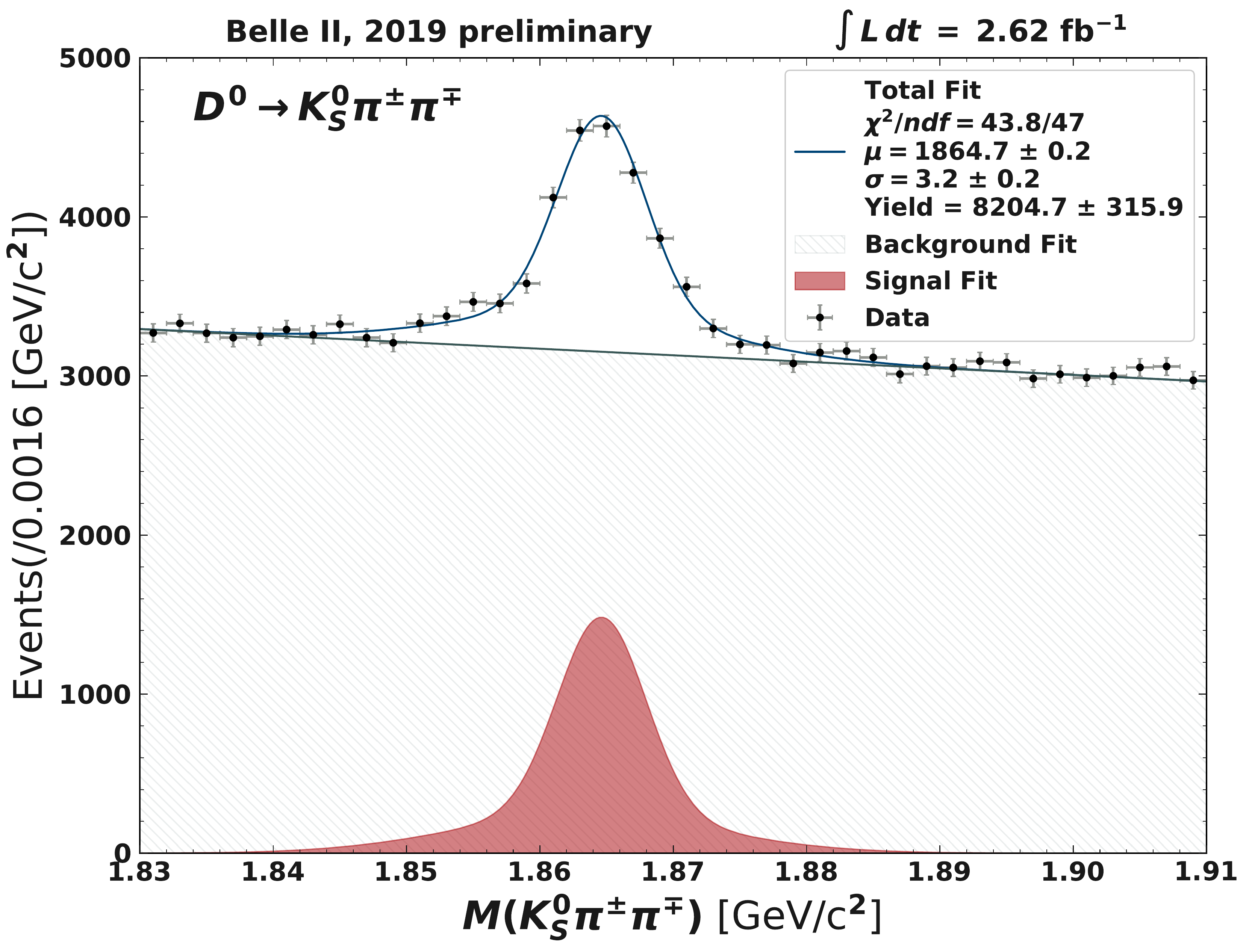}
	\includegraphics[width=.32\textwidth]{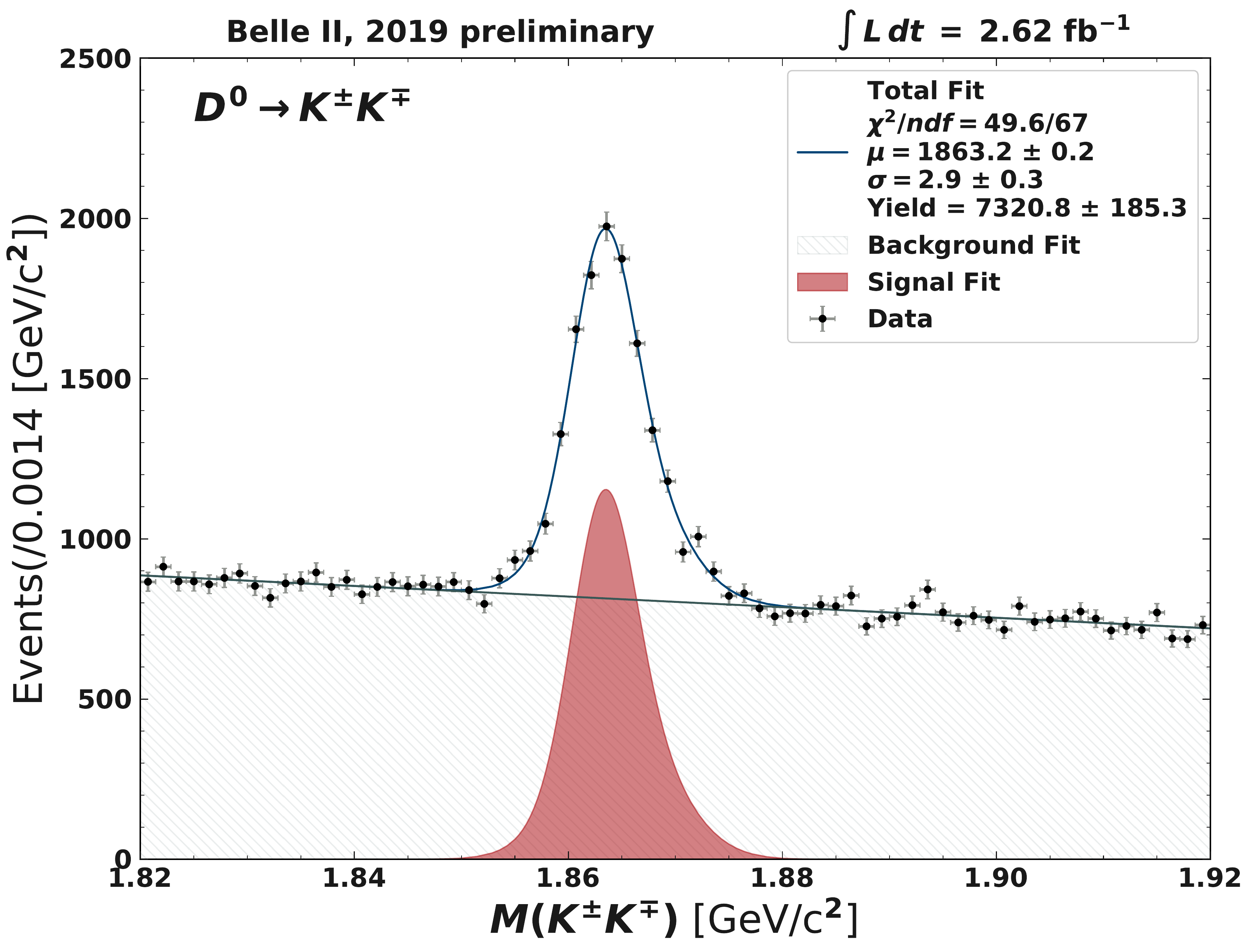}
        \caption{}
    \end{subfigure}
    ~ 
   
 \begin{subfigure}[b]{0.9\textwidth}
	\includegraphics[width=.31\textwidth]{d0_kpi_RooFitfitNopull.pdf}
	\includegraphics[width=.31\textwidth]{d0_ks2pi_RooFitfitNopull.pdf}
	\includegraphics[width=.32\textwidth]{d0_kk_RooFitfitNopull.pdf}
        \caption{}
    \end{subfigure}
	\caption{Fitted $D^{0}$ mass peaks for $0.504~$fb$^{-1}$ of 2018 comissioning data (a) and $2.62~$fb$^{-1}$ of 2019 data (b).}
\label{fig:D0_phse3}
\end{figure}
The widths of the reconstructed $D^{0}$ signals in data are consistent with the widths obtained in MC simulation.
\newpage
\section{Summary}
We present the search for the partner state of the $X(3872)$ at the $D^{*0}\bar{D}^{*0}$ threshold. For $2~$ab$^{-1}$ we expect a 
statistical significance of $6.5\sigma$ if the branching fraction of the decay $X(4014)\rightarrow D^{*0}\bar{D}^{*0}$ is assumed 
identical to one of the decay $X(3872)\rightarrow D^{0}D^{*0}$. The reconstruction of open charm meson in five different decays was tested 
with data recorded at the Belle II experiment in 2018 and 2019.

\section{Acknowledgements}

This work was supported by the BMBF (Bundesministerium für Bildung and Forschung) under grant agreement 05H15RGKBA and the Japan and Europe Network for Neutrino and Intensity Frontier Experimental Research, 
an MSCA-RISE project funded by European Union, under grant agreement 644294.

\end{document}